\documentstyle[preprint,tighten,aps,epsf,prb,eqsecnum]{revtex}
 
\begin{document}
 
\title{Dephasing time of composite fermions}
 
\author{Patrick A. Lee$^a$, Eduardo R. Mucciolo$^b$, and Henrik Smith$^c$}
 
\address{$^a$Department of Physics, Massachusetts Institute of Technology, 
Cambridge, MA 01239\\ $^b$NORDITA, Blegdamsvej 17, DK-2100 Copenhagen
{\O}, Denmark\\ $^c${\O}rsted Laboratory, H. C. {\O}rsted Institute,
Universitetsparken 5, University of Copenhagen, DK-2100 Copenhagen
{\O}, Denmark}
 
\date{April 22, 1996}
 
\maketitle
 
\begin{abstract}
We study the dephasing of fermions interacting with a fluctuating
transverse gauge field. The divergence of the imaginary part of the
fermion self energy at finite temperatures is shown to result from a
breakdown of Fermi's golden rule due to a faster than exponential
decay in time. The strong dephasing affects experiments where phase
coherence is probed. This result is used to describe the suppression
of Shubnikov-de Haas (SdH) oscillations of composite fermions
(oscillations in the conductivity near the half-filled Landau
level). We find that it is important to take into account both the
effect of dephasing and the mass renormalization. We conclude that
while it is possible to use the conventional theory to extract an
effective mass from the temperature dependence of the SdH
oscillations, the resulting effective mass differs from the $m^\ast$
of the quasiparticle in Fermi liquid theory.
\end{abstract}

\draft

 
\narrowtext

\section{Introduction}
\label{sec:intro}

It has been known for some time that fermions coupled to transverse
electromagnetic field fluctuations lead to singular corrections to the
specific heat\cite{Holstein73} and self energy.\cite{Reizer89} While
these effects are small and so far unobserved in metals, they play a
prominent role in the study of strong correlation models, such as the
$t$-$J$ model,\cite{Lee89} and more recently, in the composite fermion
description of the half-filled Landau level.\cite{Halperin93} In the
latter case, flux tubes are attached to electrons forming composite
fermions\cite{Jain89} and the mean field theory leads to a Fermi
sea. Corrections to the mean field theory are due to space- and
time-dependent density fluctuations, resulting in gauge fluctuations
which directly affect physical observables. Indeed, the fluctuations
are so strong that quasiparticles in the Landau sense may not be well
defined, and the question arises as to why the composite fermion
description appears to work so well. The question was addressed
recently by Stern and Halperin\cite{Stern95} and by Kim {\it et
al.}\cite{Kim95} In the case of Coulomb repulsion between the
electrons, the self-energy $\Sigma = \Sigma^{\prime} +
i\Sigma^{\prime\prime}$ takes the form $\Sigma^{\prime\prime}(\omega)
\sim \omega$ and $\Sigma^{\prime}(\omega) \sim \omega\ln\omega$, and
Stern and Halperin showed that the quasiparticle concept remains
marginally valid, albeit with a logarithmically divergent effective
mass. They further showed that in the quantum Boltzmann equation, the
effect of the divergent mass is cancelled by a singular Landau
function. Kim {\it et al.} treated the more singular case when the
Coulomb interaction is screened to become short ranged (by a nearby
metallic gate, for instance). They showed that even though the
quasiparticles are not well defined, a quantum Boltzmann equation can
be derived, and the cancellation between the divergent mass and the
Landau function occurs for all measurements involving a smooth
distortion of the Fermi surface. Most of the experiments, including
surface acoustic wave (SAW) resonances,\cite{Willett93} magnetic
focusing,\cite{Goldman94} and anti-dot resonances\cite{Kang93} belong
to this category. It is expected that the divergent mass will show up
only in the activation gap of the effective Landau levels of the
composite fermions.

The issue of the effective mass is related to the real part of the
self energy. In this paper we address the issue of how the imaginary
part of the self energy will affect experiments. The experiments we
have in mind are those which require phase coherence of the composite
fermions, such as mesoscopic phenomena and quantum oscillations [de
Haas-van Alphen (dHvA) and SdH effects]. Such phenomena are outside
the purview of the quantum Boltzmann equation. The difficulty which we
immediately encounter is that, at finite temperatures,
$\Sigma^{\prime\prime}(T)$ is infinite.\cite{Lee90} This is because
the thermal factor $k_BT/\omega$ for the soft gauge fluctuations gives
rise to a further singularity at small $\omega$ which has no obvious
cutoff. Here we first analyze this problem and show that the
difficulty results from a breakdown of Fermi's golden rule
(Sec.~\ref{sec:breakrule}). We then use a semiclassical approach to
discuss the suppression of the SdH oscillations due to the dephasing
of composite fermions (Sec.~\ref{sec:ShubdeHa}). The effect of mass
renormalization is considered in Sec.~\ref{sec:massrenorm}. Our
theoretical results are compared with experimental data in
Sec.~\ref{sec:compare} and conclusions are drawn in
Sec.~\ref{sec:conclude}.

\section{Breakdown of Fermi's golden rule}
\label{sec:breakrule}

We begin by treating the Green's function of a particle in a space-
and time-dependent gauge field ${\bf a}({\bf r},t)$ and potential
$a_0({\bf r},t)$ using the semiclassical (Gorkov) approximation
\begin{equation}
G({\bf r},t) = G_0({\bf r},t) e^{i\phi(t)},
\label{eq:gorkov}
\end{equation}
where $G_0$ is the free fermion Green's function and
\begin{equation}
\phi(t) = \int_0^t \left[ {\bf a} \biglb( {\bf r}(t^\prime),t^\prime
\bigrb) \cdot {\bf v}_F + a_0 \biglb( {\bf r}(t^\prime),t^\prime
\bigrb) \right]\ dt^\prime
\label{eq:phase}
\end{equation}
(our units are such that $\hbar=c=k_B=1$). We assume that the particle
travels in a straight line with velocity ${\bf v}_F$, so that ${\bf
r}(t^\prime) = {\bf r} t^\prime/t = {\bf v}_F t^\prime$. Since the
gauge fluctuations scatter mainly in the forward direction, the
velocity is affected only on a long time scale (the transport time) so
that this assumption is justified. Equation~(\ref{eq:phase}) has the
property that under a gauge transformation, ${\bf a} \rightarrow {\bf
a} + \nabla \Lambda$, $a_0 \rightarrow a_0 + \dot{\Lambda}$, $\phi(t)
\rightarrow \phi(t) + \Lambda\biglb( {\bf r}(t),t \bigrb) -
\Lambda\biglb( {\bf r}(0),0 \bigrb)$, as is required for the gauge
transformation of $G({\bf r},t)$ given by Eq.~(\ref{eq:gorkov}). We
shall work in the transverse gauge, $\nabla\cdot {\bf a} = 0$, where
it can be shown that the contribution from $a_0$ fluctuations are
unimportant, and will be ignored from here on. The fluctuations of
${\bf a}$ are Gaussian and controlled by the correlation function
\begin{equation}
\langle \tilde{a}_\alpha({\bf q},\omega) \tilde{a}_\beta(-{\bf
q},-\omega) \rangle = \coth\left(\frac{\omega}{2T}\right) \left(
\delta_{\alpha\beta} - \frac{q_\alpha q_\beta}{q^2} \right) {\rm Im}
D_{11}(q,\omega) \;,
\label{eq:Gaussaver}
\end{equation}
where, for $\omega<qv_F$ and $q\ll k_F$, the retarded transverse gauge
propagator is given by\cite{Halperin93}
\begin{equation}
D_{11}(q,\omega) = \frac{1}{i\omega\gamma_q - q^2
\tilde{\chi}^\prime(q)} \;,
\label{eq:propagator}
\end{equation}
with
\begin{eqnarray}
\gamma_q^{-1} & = & \frac{2\pi q}{k_F} \quad \mbox{and} \quad
\tilde{\chi}^\prime(q) = \tilde{\chi}_0^\prime +
\frac{\tilde{v}(q)}{(4\pi)^2} \nonumber \;.
\end{eqnarray}
In the expressions above $\tilde{\chi}_0^\prime$ represents the
diamagnetic susceptibility of the composite fermions at the
half-filled state, $k_F=mv_F$ is the Fermi momentum, and
$\tilde{v}(q)$ is the Fourier transform of the two-body interaction
potential $v(r)$. We refer to $m$ as the mean-field mass of the
composite fermion which arises from the interaction energy and differs
substantially from the band mass of an electron. For convenience, we
shall simply write
\begin{equation}
D_{11}(q,\omega) = \frac{(2\pi q/k_F)}{i\omega - C_\eta q^{\eta+1}} \;.
\label{eq:newdefD}
\end{equation}
In the Coulomb case $v(r)=e^2/(\epsilon r)$, where $\epsilon$ is the
medium dielectric constant, therefore $\eta=1$ and $C_1 =
e^2/(4\epsilon k_F)$. For the short-range, $\delta$-function
interaction $\eta=2$ and $C_2 = 2\pi \tilde{\chi}^\prime(0)/k_F$.

From Eq.~(\ref{eq:gorkov}) we see that the dephasing of the fermion is
given by the factor $\langle \exp[ i \phi(t)] \rangle =
e^{-F_\eta(t)}$, where\cite{Lee90}
\begin{eqnarray}
F_\eta(t) & = & \frac{1}{2} \langle \phi^2(t) \rangle \nonumber \\ & =
& -\int \frac{d^2q}{(2\pi)^2} \int_0^\infty \frac{d\omega}{\pi} |{\bf
v}_F \times \hat{{\bf q}}|^2 \coth \left( \frac{\omega}{2T} \right)
{\rm Im} D_{11}(q,\omega) \left\{ \frac{1 - \cos[({\bf v}_F\cdot {\bf
q} - \omega)t]}{({\bf v}_F \cdot {\bf q} - \omega)^2} \right\}.
\label{eq:defFfunction}
\end{eqnarray}
The term appearing in $\{\ldots\}$ is well known in elementary quantum
mechanics. When $t$ is large, it is a strongly peaked function of
$({\bf v}_F\cdot {\bf q} - \omega)$ with height $\sim t^2$ and width
$\sim t^{-1}$ and is usually approximated by $2\pi t \delta({\bf
v}_F\cdot {\bf q} -\omega)$. The $\delta$-function is recognized as
enforcing energy conservation (after ignoring the $q^2/2m$ term) and
$F_\eta(t)$ takes the form $t\Sigma^{\prime\prime}$, where
$\Sigma^{\prime\prime}(T)$ is the usual formula for the imaginary part
of the self energy. The exponential decay $\exp[-F(t)]$ then leads to
the interpretation of $\Sigma^{\prime\prime}$ as the inverse
lifetime. This is the standard derivation of Fermi's golden
rule. However, for our problem $\Sigma^{\prime\prime}(T)$ is infra-red
divergent for both $\eta=$1 and 2. On the other hand, $F_\eta(t)$ is
finite for all $t$ when $\eta<2$. [$F_2(t)$ is logarithmically
divergent, as is easily seen by a small $t$ expansion.\cite{Lee90}] We
therefore directly evaluate the integral in
Eq.~(\ref{eq:defFfunction}) by the following steps. In order to
eliminate the singular denominator in Eq.~(\ref{eq:defFfunction}) we
first consider the second derivative $\ddot{F}_\eta(t)\equiv d^2
F_\eta(t) /dt^2$. Then we perform the $q$ integration (details are
shown in Appendix A). We find (for $\eta=1$)
\begin{equation}
\ddot{F}_1(t) = \left( \frac{v_F^2}{2k_F C_1^{3/2}} \right)
\int_0^\infty d\omega \sqrt{\omega} \coth \left( \frac{\omega}{2T}
\right) \cos(\omega t)\ f\left(v_F t \sqrt{\frac{\omega}{C_1}} \right)
\;,
\label{eq:Fdoubleprime}
\end{equation}
where $C_1$ is defined below Eq.~(\ref{eq:newdefD}) and
\begin{equation}
f(x) = \frac{2}{\pi} \int_0^{\pi/2} d\theta \cos^2\theta\ \exp
\left(-\frac{x\sin\theta}{\sqrt{2}} \right) \cos \left(
\frac{x\sin\theta}{\sqrt{2}} + \frac{\pi}{4} \right) \;.
\label{eq:f_func}
\end{equation}
We conclude that $\ddot{F}_1(t)$ has two regimes. For $t\ll t_T$,
where $t_T\equiv (1/v_F)\sqrt{C_1/T}$, it is given by $1/(2^{3/2} E_F
t_T^3)[1-2^{3/2}t/(3\pi t_T)]$, whereas for $t\gg t_T$, it is $1/(2
E_F t_T^2 t)$. Upon integration, we find that
\begin{eqnarray}
F_1(t) \simeq \left\{\begin{array}{lr} \frac{1}{2^{3/2}} \frac{1}{2
E_F t_T} \left( \frac{t}{t_T} \right)^2 + O(t/t_T)^3 & t \ll
t_T \\ \frac{1}{2 E_F t_T} \left( \frac{t}{t_T} \right) \ln \left(
\frac{t}{t_T} \right) & t \gg t_T 
\end{array}\right. \;,
\label{eq:FCoulomb}
\end{eqnarray}
where $E_F=k_F^2/2m$ is the Fermi energy. Thus we see that the
dephasing of the fermion is faster than exponential in time, which
explains why the self energy $\Sigma^{\prime\prime}$ is infinite,
because this amounts to force-fitting to an exponential decay. The
technical reason is that the replacement of the $\{\ldots\}$ factor in
Eq.~(\ref{eq:defFfunction}) by a $\delta$-function is invalid if the
rest of the integrand is singular in the small $q$ and $\omega$
limit. This explains the breakdown of Fermi's golden rule. Similar
considerations show that the long-time behavior for $F_\eta(t)$ is
proportional to $t^{\eta}/(\eta-1)$ for $1<\eta<2$.

\section{Suppression of quantum oscillations}
\label{sec:ShubdeHa}

The Green's function in Eq.~(\ref{eq:gorkov}) is not gauge invariant
and is not a directly measurable quantity. Nevertheless, the strong
dephasing found in the previous section does manifest itself
physically by suppressing quantum oscillations. Perhaps the clearest
example is the strong damping of oscillations in the longitudinal
conductivity $\sigma_{\rm xx}(B)$ near the half-filled state. The
damping is mainly due to the thermal fluctuations of the field ${\bf
a}({\bf r},t)$. This phenomenon is essentially the SdH effect for
composite fermions.

The approach outlined in the previous section will now be used to
discuss physical measurements such as the SdH effect. We shall employ
the formulation in terms of semiclassical paths, which can describe
quantum oscillations in unmodulated electron gases, as well as in
anti-dot
arrays.\cite{Reichl92,Aronov95,antidots,Hackenbroich95,Mirlin95} The
oscillatory part of the conductivity measured in a SdH experiment,
$\sigma^{\rm osc}_{\rm xx}$, comes mostly from the interference
between repetitions of the same cyclotron trajectory [represented by
the off-diagonal term in Eq.~(15) of
Ref.~\onlinecite{Hackenbroich95}]. As a consequence of that,
$\sigma^{\rm osc}_{\rm xx}$ can be expressed in a form very similar to
the oscillations in the density of states, which are proportional to
the integral
\begin{equation}
\int_{-\infty}^{\infty} d\varepsilon\ n_F(\varepsilon)\ {\rm Re}
\sum_{p=1}^\infty (-1)^p \left\langle e^{iS_p(\varepsilon) +
i\varphi_0} \right\rangle \;,
\label{eq:traceform}
\end{equation}
where $n_F(\varepsilon)=(e^{\varepsilon/T}+1)^{-1}$ and $\varphi_0$ is
a constant. We call $S_p(\varepsilon)$ the action of a classical path
with energy $\varepsilon$ which traverses a cyclotron orbit $p$
times. In the presence of a fluctuating gauge field we have
\begin{equation}
\left\langle e^{iS_p(\varepsilon)} \right\rangle = e^{i2\pi
p\varepsilon/\omega_c} \left\langle e^{i \oint {\bf a}\biglb( {\bf
r}_p(t), t \bigrb)\cdot {\bf v}_F dt} \right\rangle \;,
\label{eq:phasefactor}
\end{equation}
where ${\bf r}_p(t)$ is the classical orbit and $\omega_c=eB/m$ is the
cyclotron frequency. ($B$ denotes the external magnetic field felt by
the composite fermions.) More generally, the phase factor in
Eq.~(\ref{eq:phasefactor}) should include the $a_0$ term, as in
Eq.~({\ref{eq:phase}), so that under a gauge transformation
$S_p(\varepsilon)$ will be shifted by $\chi \biglb( {\bf r}(t),t
\bigrb) - \chi \biglb( {\bf r}(0),0 \bigrb)$. In a gauge-invariant
formulation of the SdH oscillations, this phase factor should be
cancelled by a corresponding gauge transformation of $\varphi_0$. For
our purposes it suffices to note that a gauge transformation
introduces only a fixed phase shift in $\left\langle
e^{iS_p(\varepsilon)} \right\rangle$, so that as far as the damping
factor is concerned, it can be evaluated in any gauge. In particular,
in the transverse gauge the dominant contributions come from the
transverse gauge fluctuations, and Eq.~(\ref{eq:phasefactor}) can be
used. In any other gauge, $a_0$ fluctuations must be included, which
complicates the calculation.
 
There is one further complication we must overcome before we can
proceed with the evaluation of Eq.~(\ref{eq:phasefactor}). If we
evaluate Eq.~(\ref{eq:phasefactor}) using Eq.~(\ref{eq:defFfunction}),
it was noted before\cite{Lee89,Lee90} that $F_\eta(t)$ diverges
logarithmically for any path in the short-range case ($\eta=2$). This
is because the fluctuations in ${\bf a}$ can get arbitrarily large,
even though the fluctuations in ${\bf h} = \nabla \times {\bf a}$ are
bounded. The divergence is avoided for closed paths in the case when
the gauge field ${\bf a}$ is not explicitly time dependent. In this
case, $S_p(\varepsilon)$ can be written explicitly in terms of the
fluctuations of the field ${\bf h}$, which are bounded. On the other
hand, for dynamical fluctuations with frequency $\omega>\omega_c/p$,
the flux enclosed by the orbit has changed by the time the particle
completes $p$ cycles, and $S_p$ can no longer be written in terms of
${\bf h}( {\bf r},t)$. For these paths we can use the Gorkov
approximation discussed earlier. Thus our strategy is the
following. We divide the frequency spectrum for the gauge fluctuations
into two parts: the part with $\omega<\omega_c/p$ will be considered
static and treated in an explicitly gauge invariant way in terms of
the flux fluctuations; the remaining part will be treated in the way
discussed earlier, except that now a low-frequency cutoff $\omega_c/p$
is introduced. This part of the gauge fluctuation is the same for
either closed or open paths and our discussion in the previous section
is applicable.

The dephasing due to static gauge fluctuations was first discussed by
Aronov and coworkers\cite{Aronov95} for $\delta$-correlated flux
fluctuations and their treatment can be easily extended to the more
general case presented here. It yields a dephasing factor $e^{-W_s}$,
where
\begin{eqnarray}
W_s & = & \frac{1}{2} \left\langle \left(\oint {\bf a} \cdot d{\bf l}
\right)^2 \right\rangle \nonumber \\ & = & \frac{p^2}{2} \int d^2r_1
\int d^2r_2\ \langle h({\bf r}_1)h({\bf r}_2) \rangle \;,
\label{eq:static}
\end{eqnarray}
with $\langle | h({\bf q},\omega) |^2 \rangle = q^2 \langle | {\bf
a}({\bf q},\omega) |^2 \rangle$. (The line integration is performed
over a cyclotron trajectory of winding number $p$ and the surface
integrations encompass a circle of radius $R_c=v_F/\omega_c$.) Note
that the $q^2$ factor gives additional convergence at small $q$,
making all integrals finite. For the Coulomb case we find that (see
Appendix B)
\begin{equation}
W_s = \frac{\pi p^2 T}{m C_1 \omega_c} \ln \left( \frac{E_F}{p
\omega_c} \right)
\label{eq:statC}
\end{equation}
for $E_F\gg\omega_c$.

For the dynamical part, we proceed as in Sec.~\ref{sec:breakrule}, the
only new feature being that the $\omega_c$ cutoff introduces a new
time scale $t_{\omega_c} \equiv (1/v_F)\sqrt{C_1/\omega_c}$. We find
that $\ddot{F}_1(t)$ decreases as $t^{-3}$ for $t\gg t_{\omega_c}$
(see Appendix A). This decrease is fast enough so that $F_1(t)$ is now
linear in $t$ for $t\gg t_{\omega_c}$. From now on we shall restrict
ourselves to $T\gg\omega_c$, so that only the $p=1$ orbit is
important. We are interested in evaluating $F_1(t)$ at
$t=2\pi/\omega_c$, which is the linear regime. We conclude that the
spectrum for $\omega>\omega_c$ contributes to a dephasing factor
$e^{-W_d}$, where
\begin{equation}
W_d = \frac{\pi T}{m C_1 \omega_c} \ln \left( \frac{T}{\omega_c}
\right) \;.
\label{eq:dynamical}
\end{equation}
Note that $W_d$ is smaller than $W_s$ if $T<E_F$. 

For short-range repulsion ($\eta=2$) we can carry out a similar
analysis. We find that
\begin{eqnarray}
F_2(t) \simeq \left\{\begin{array}{lc} \frac{v_F T t^2}{12 m C_2} \ln
\left( \frac{T}{\omega_c} \right), & t \ll \frac{1}{v_F} \left(
\frac{C_2}{T} \right)^{1/3} \\ \frac{v_F T t^2}{12 m C_2} \ln \left(
\frac{C_2}{\omega_c v_F^3 t^3} \right), & \frac{1}{v_F} \left(
\frac{C_2}{T} \right)^{1/3} \ll t \ll \frac{1}{v_F} \left(
\frac{C_2}{\omega_c} \right)^{1/3} \\ \frac{2T t}{\sqrt{3}\pi m
C_2^{2/3} \omega_c^{1/3}}, & t \gg \frac{1}{v_F} \left(
\frac{C_2}{\omega_c} \right)^{1/3}
\end{array}\right. \;.
\label{eq:Fshortrange}
\end{eqnarray}
Consequently, the spectrum for $\omega>\omega_c$ contributes a
dephasing factor with argument
\begin{equation}
W_d = \frac{4\pi T}{\sqrt{3} m C_2^{2/3} \omega_c^{4/3}} \;.
\label{eq:Wd_short}
\end{equation}
This is to be compared with
\begin{equation}
W_s = \frac{\pi^2 v_F T}{2m C_2 \omega_c^2} \;,
\label{eq:Ws_short}
\end{equation}
a result first obtained by Mirlin {\it et al.}\cite{Mirlin95} We note
that the introduction of an upper cutoff $\omega_c$ in the $\omega$
integral did not change the result for $W_s$. Comparing
Eq.~(\ref{eq:Ws_short}) with (\ref{eq:Wd_short}) we conclude that for
short-range interactions, quasi-static fluctuations dominate the
dephasing, so that $W_d$ can be neglected.

We note in passing that in carrying out the integration, we find that
the factor $\cos(\omega t)$ in Eq.~(\ref{eq:Fdoubleprime}) can be set
to unity for all values of $t$. This means that our result would be
the same if we had treated the ${\bf a}({\bf r},t)$ field as static
from the beginning, and written the phase factor as $\int {\bf a({\bf
r})} \cdot d{\bf l}$. This shows that the suppression factor is purely
geometrical, depending only on the orbit circumference and not on the
transit time. Thus any renormalization of the velocity of the particle
will not affect our estimate. Furthermore, in the Coulomb case
($\eta=1$), if we do not introduce the separation into static and
dynamical part, and treat the gauge field as purely static, we see
(from Appendix B) that the suppression factor is correctly given as
$W_s + W_d$. Alternatively, we could also treat all fluctuations as
``dynamical'', using the Gorkov approximation. The suppression factor
would then be given by $F_1(t=2\pi/\omega_c)$, where $F_1$ is given by
Eq.~(\ref{eq:FCoulomb}). We can see that the result $W_s + W_d$ is
correctly reproduced.

However, for the general $\eta$, it is necessary to separate the
spectrum of the gauge fluctuations into high and low frequency
components. Nevertheless, the final answer is that the dominant
dephasing factor is correctly given by the static approximation
evaluated in terms of the flux fluctuations, as was done in
Refs.~\onlinecite{Aronov95} and \onlinecite{Mirlin95}. Thus, the
results of this section can be viewed as a justification of this
intuitively appealing approach. With the understanding we have gained,
we can now proceed to include the effect of mass renormalization,
where dynamical gauge fluctuations are essential.

\section{Mass renormalization}
\label{sec:massrenorm}

The semiclassical method yields a dephasing time for the composite
fermion, but it cannot address the issue of the renormalization of the
mass (or the energy gap) due to virtual dynamical gauge
fluctuations. In the case of SdH oscillations, thermal broadening of
the Landau levels plays an important role and it is necessary to
include mass renormalization on the same footing as dephasing. To do
this we return to a many-body diagrammatic treatment. For the Coulomb
case, the work of Stern and Halperin suggests that Fermi liquid theory
remains valid. We shall therefore use the standard treatment, such as
that given by Engelsberg and Simpson\cite{Engelsberg70} for the dHvA
magnetization in the presence of a strong electron-phonon
coupling. From semiclassical arguments we expect similar
density-of-state oscillations to appear in the SdH effect as
well.\cite{Mirlin95} For spinless particles moving in two dimensions,
the oscillatory part of the dHvA magnetization is given
by\cite{Engelsberg70}
\begin{eqnarray}
\frac{M_{\rm osc}(B)}{\Omega} & = & \frac{m E_F}{\pi B} \mbox{Re}
\left\{ \sum_{k=1}^\infty (-1)^k e^{2\pi ik E_F/\omega_c}
\int_{-\infty}^\infty d\varepsilon\ n_F(\varepsilon)\ e^{(2\pi
ik/\omega_c)[\varepsilon - \Sigma^\prime(\varepsilon)
-i\Sigma^{\prime\prime}(\varepsilon)]} \right\} \label{eq:dHvA1} \\ &
= & \frac{2 m T E_F}{B} \sum_{k=1}^\infty (-1)^{k} \sin \left( \frac{2
\pi k E_F}{\omega_c} \right) \sum_{n=0}^\infty e^{-(2\pi k/\omega_c) [
\omega_n + i\Sigma(i\omega_n) ]} \;, \label{eq:dHvA2}
\end{eqnarray}
where $\Omega$ is the sample area and $\omega_n=\pi T(2n+1)$, while
$\Sigma(i\omega_n) = \Sigma^\prime(i\omega_n) +
i\Sigma^{\prime\prime}(i\omega_n)$. Equation~(\ref{eq:dHvA1}) is
suggestive of the semiclassical formulas of the previous section, with
$2\pi k\Sigma^{\prime\prime}/\omega_c$ playing the role of the
dephasing factor. In addition, we now have a modification of the
energy due to $\Sigma^\prime$. Equations~(\ref{eq:dHvA1}) and
(\ref{eq:dHvA2}) cannot be used directly because $\Sigma$ is divergent
at finite $T$. We now adopt the same strategy as before and separate
the frequency of the gauge fluctuations into $\omega<\omega_c$ and
$\omega>\omega_c$. The $\omega<\omega_c$ part we treat as static,
while the $\omega>\omega_c$ part is treated using Eq.~(\ref{eq:dHvA1})
with a low-energy cutoff. Therefore, to lowest order in the gauge
field, we use the following formula for the (retarded) self energy:
\begin{eqnarray}
\Sigma({\bf k},\varepsilon) & = & -\int \frac{d^2q}{(2\pi)^2}
\int_{\omega_c}^{qv_F} \frac{d\omega}{\pi} \frac{|{\bf k}\times
\hat{\bf q}|^2}{m^2} \mbox{Im}D_{11}(q,\omega) \left[
\frac{1+n_B(\omega)-n_F(\xi_{k+q})}{\varepsilon + i0^+ - \xi_{k+q}
-\omega} \right.  \nonumber \\ & & \left.  +
\frac{n_B(\omega)+n_F(\xi_{k+q})}{\varepsilon + i0^+ - \xi_{k+q}
+\omega} \right] \;,
\label{eq:diagram}
\end{eqnarray}
where $n_B(\varepsilon)=(e^{\varepsilon/T}-1)^{-1}$ and $\xi_k =
k^2/(2m)-E_F$.

For Coulomb interactions and $\varepsilon\ll T$, we find that
\begin{equation}
\Sigma^\prime(k_F,\varepsilon) = -\frac{\varepsilon}{2\pi m C_1} \ln
\left( \frac{E_F}{T} \right)
\label{eq:relse}
\end{equation}
and
\begin{equation}
\Sigma^{\prime\prime}(k_F,\varepsilon) = -\frac{T}{2 m C_1} \ln \left(
\frac{T}{\omega_c} \right) \;.
\label{eq:imagse}
\end{equation}
The contribution coming from $\Sigma^{\prime\prime}$ is identical to
the semiclassical result given by Eq.~(\ref{eq:dynamical}). On the
other hand, the real part $\Sigma^\prime$ leads to a renormalization
of the mass given by\cite{Stern95}
\begin{equation}
m^{\ast} = m \left[ 1 + \frac{1}{2\pi m C_1} \ln \left( \frac{E_F}{T}
\right) \right] \;,
\label{eq:sternhalp}
\end{equation}
which, as stated before, cannot be obtained in a semiclassical
treatment. Upon performing the $\epsilon$ integration in
Eq.~(\ref{eq:dHvA1}) in the standard way, we find for $T>\omega_c$
that it is sufficient to keep the $k=1$ and $n=0$ term and the dHvA
oscillations are suppressed by the factor
\begin{equation}
A_{\rm dynamic} = \exp \left( -\frac{2\pi^2 T}{\omega_c^\ast} \right)
\exp \left[ -\frac{2\pi \Sigma^{\prime\prime}(T)}{\omega_c} \right],
\label{eq:suppfac}
\end{equation}
where $\omega_c^\ast=eB/m^\ast$. The first term in
Eq.~(\ref{eq:suppfac}) comes from the thermal smearing of the
renormalized Landau levels, while the second term comes from the
dephasing of the composite fermion. In metals this second term is
usually ignored relatively to the first one because the electron
scattering rate scales as $T^2$. Here the two terms are comparable. In
fact, while $\omega_c^\ast$ and $\Sigma^{\prime\prime}$ separately
have a $\ln T$ dependence, we find that this is cancelled when we
combine the two factors together, yielding
\begin{equation}
A_{\rm dynamic} = \exp \left\{ - \frac{2\pi^2 T }{\omega_c} \left[ 1 +
\frac{1}{2\pi m C_1} \ln \left( \frac{E_F}{\omega_c} \right) \right]
\right\}.
\label{eq:suppres}
\end{equation}
Remarkably, as the temperature is raised, the dephasing of the fermion
is exactly balanced by the reduction of the mass enhancement. This
cancellation of the $T$ dependence also occurs in the electron-phonon
problem, as $T$ goes from below to above the Debye temperature, as was
shown in Ref.~\onlinecite{Engelsberg70} using Eq.~(\ref{eq:dHvA2}). We
can check that in our case Eq.~(\ref{eq:suppres}) also follows from
Eq.~(\ref{eq:dHvA2}) because
\begin{equation}
\Sigma(i\omega_n) = -\frac{i T}{2 m C_1} \ln \left(
\frac{E_F}{\omega_c} \right) \;.
\label{eq:anconse}
\end{equation}
Finally, to obtain the total suppression factor, we have to include
the contribution from $\omega<\omega_c$, which is given by
$e^{-W_s}$. For the case of long-range Coulomb interactions, combining
all contributions, we find that the suppression of the amplitude of
the conductivity oscillations is given by
\begin{equation}
\delta\sigma^{\rm osc}_{\rm xx}(B) \propto \exp \left\{ -\frac{2\pi^2
T}{\omega_c} \left[ 1 + \frac{1}{\pi m C_1} \ln \left(
\frac{E_F}{\omega_c} \right) \right] \right\} \;.
\label{eq:sigmaosc}
\end{equation}
This expression implies that it is possible to analyze the $T$
dependence of the SdH amplitude using the conventional theory, as was
done in Ref.~\onlinecite{Du93}, and obtain a temperature-independent
effective mass $m_{\rm eff}$ from the slope of $\ln(\delta\sigma^{\rm
osc}_{\rm xx})$ versus $T$. This effective mass is predicted to show a
logarithmic divergence near $B=0$, i.e.,
\begin{equation}
m_{\rm eff}(B) = m \left[ 1 + \frac{1}{\pi m C_1} \ln \left(
\frac{E_F}{\omega_c} \right) \right] \;.
\label{eq:massenh}
\end{equation}
Notice that the prefactor of the logarithm is twice that given by
Eq.~(\ref{eq:suppres}), which is the same as $m^\ast$ given by
Eq.~(\ref{eq:sternhalp}) with $T$ replaced by $\omega_c$ as a
cutoff. Thus one cannot identify $m_{\rm eff}$ with the $m^\ast$ from
Fermi liquid theory. The difference comes from the essential
contribution to the dephasing due to the static component of the gauge
fluctuation.

A similar analysis can be carried out for short-range
interactions. Note, however, that the use of results obtained in
Ref.\onlinecite{Engelsberg70} is much more questionable in this case,
since Fermi liquid theory does not apply for the short-range
interactions. We find that, close to the Fermi surface,
\begin{equation}
\Sigma^\prime(k_F,\varepsilon) \approx - \frac{\alpha \varepsilon}{m
C_2^{2/3} T^{1/3}}
\label{eq:rpsesr}
\end{equation}
and
\begin{equation}
\Sigma^{\prime\prime}(k_F,\varepsilon) = - \frac{2T}{\sqrt{3} m
C_2^{2/3} \omega_c^{1/3}} \;,
\label{eq:imagsesr}
\end{equation}
where $\alpha$ is a number of order $10^{-1}$. The contribution of the
mass renormalization to the damping factor is now negligible in
comparison with $\Sigma^{\prime\prime}/\omega_c$. The latter accounts
exactly for the dynamical term [Eq.~(\ref{eq:Wd_short})]. Hence the
damping of the conductivity oscillations is dominated by the static
term of Eq.~(\ref{eq:Ws_short}). It leads to an effective mass of the
form $m_{\rm eff}(B) = m [1 + (2 m k_F C_2)^{-1} (E_F/\omega_c)]$.

\section{Comparison with experiments}
\label{sec:compare}

Several recent
experiments\cite{Du93,Leadley94,Manoharam94,Du94,Coleridge95} have
measured the oscillations in the longitudinal conductance near
half-filling. The data were fitted using the conventional theory of
SdH oscillations for non-interacting electrons, by allowing their
effective mass $m_{\rm eff}(B)$ to depend on magnetic field. With the
exception of Ref.\onlinecite{Leadley94}, these studies indicate that
$m_{\rm eff}(B)$ is enhanced as the magnetic field felt by the
composite fermions $B = B_{\rm ext} - B_{1/2}$ approaches zero.

In the case of Coulomb interactions ($\eta=1$), our results are
consistent with a logarithmically divergent effective mass. For
$1<\eta\le 2$ we find that $m_{\rm eff}(B)$ should diverge as
$B^{1-\eta}$. This enhancement, however, is not as strong as that
observed in the experiments. To quantify this, we have attempted to
fit Eq.~(\ref{eq:massenh}) to the high effective-field part of the
data obtained by Du {\it et al.}\cite{Du94} Following
Ref.~\onlinecite{Halperin93}, we have estimated the mean-field mass by
assuming that it is related to the strength of the Coulomb interaction
through $m^{-1} = 4\beta C_1$, where $C_1$ is given below
Eq.~(\ref{eq:newdefD}) and $\beta$ is a numerical constant. Our fit
[see Fig.~\ref{fig:datafit}] indicates that $\beta\approx 0.15$,
yielding $m\approx 0.76m_0$, with $m_0$ being the free electron
mass. The values are reasonably close to those obtained from numerical
simulations of finite-size systems, such as those carried out recently
by Morf and d'Ambrumenil.\cite{Morf95} These authors found that
$\beta\approx 0.20$, which implies $m\approx 0.55m_0$.

The reason for the discrepancy at small $B$ may be due to the fact
that disorder has not been taken into account in our treatment. Close
to half-filling, where the apparent increase in the effective mass is
large, the activation gap approaches zero and it is likely that
disorder effects are playing a dominant role.

\section{Summary and conclusions}
\label{sec:conclude}

In this work we have demonstrated how the divergence, due to gauge
fluctuations, of the imaginary part of the fermion self energy is
related to the non-exponential time dependence of the dephasing
factor. We have calculated the dephasing factor for both Coulomb and
short-range interactions by separating the spectrum of gauge
fluctuations into static and dynamic parts. We have considered the
effect of dephasing and mass renormalization on the oscillating
component of the density of states in a magnetic field by relating our
approach to that of Engelsberg and Simpson for the case of
electron-phonon interactions. Our calculated oscillations in the
density of states are compared to the observed Shubnikov-de Haas
oscillations in order to extract an effective mass of the composite
fermions. While we can account for part of the apparent increase in
the mass as the $\nu=1/2$ state is approached, we are not able to
explain the strong divergence observed in the immediate vicinity of
the $\nu=1/2$ state. This may be due to the effect of disorder, which
is not included in the present description.

\section{Acknowledgements}

We would like to thank B. Altshuler, D. Khveshchenko, A. Mirlin, and
P. W\"olfle for stimulating discussions. P.A.L. is thankful for the
hospitality of NORDITA, where this work was initiated. He also
acknowledges the support of a Guggenheim Fellowship and the support
from NSF under Grant No. DMR-9523361.

\appendix
\section{}
\label{sec:appdA}

In this appendix we derive Eq.~(\ref{eq:FCoulomb}) in detail and point
out the important steps in the derivation of
Eq.~(\ref{eq:Fshortrange}). First we notice that upon differentiating
twice with respect to time, Eq.~(\ref{eq:defFfunction}) becomes
\begin{equation}
\ddot{F}_\eta(t) = -\int \frac{d^2q}{(2\pi)^2} \int_0^\infty
\frac{d\omega}{\pi} |{\bf v}_F \times \hat{{\bf q}}|^2 \coth \left(
\frac{\omega}{2T} \right) {\rm Im} D_{11}(q,\omega) \cos({\bf
v}_F\cdot {\bf q}t) \cos(\omega t),
\end{equation}
since terms which are odd in ${\bf q}$ do not contribute. Introducing
the gauge field propagator from Eq.~(\ref{eq:newdefD}), we then have
\begin{equation} 
\ddot{F}_\eta(t) = \frac{v_F^2}{\pi k_F} \int_0^\infty d\omega\
\omega \coth \left( \frac{\omega}{2T} \right) \cos (\omega t)
\int_0^{2\pi} \frac{d\varphi}{2\pi} \sin^2\varphi \int_0^{2k_F}
\frac{dq\ q^2 \cos(qv_F t \cos\varphi)}{\omega^2 + C_\eta^2
q^{2(\eta+1)}} \;,
\label{eq:fppexp}
\end{equation}
where $\varphi$ is the angle between ${\bf v_F}$ and ${\bf q}$. To
proceed, we notice that the $q$ integration can be carried out by
contour in the complex plane when $\omega\ll C_\eta k_F^{\eta+1}$,
which is satisfied provided that $\omega\ll E_F$. Specializing our
results to $\eta=$ 1 and 2, we need to calculate two integrals,
namely,
\begin{equation}
\int_0^\infty dz\ \frac{z^2 \cos(zx)}{z^4 + 1} = 2\pi
e^{-x/\sqrt{2}} \cos \left( \frac{x}{\sqrt{2}} +
\frac{\pi}{4} \right)
\label{eq:int1}
\end{equation}
and
\begin{equation}
\int_0^\infty dz\ \frac{z^2 \cos(zx) }{z^6 + 1} = \frac{4\pi}{3}
\left[ e^{-x/2} \cos \left( \frac{x \sqrt{3}}{2} \right) -
\frac{1}{2}e^{-x} \right] ,
\end{equation}
with $x>0$. Introducing Eq.~(\ref{eq:int1}) into
Eq.~(\ref{eq:fppexp}), we arrive at
Eq.~(\ref{eq:Fdoubleprime}). Analogously, for the short-range case we
find that
\begin{equation}
\ddot{F}_2(t) = \frac{v_F^2}{3 k_F C_2} \int_0^\infty d\omega\
\coth \left( \frac{\omega}{2T} \right) \cos (\omega t) g\Biglb( v_F
t (\omega/C_2)^{1/3} \Bigrb),
\end{equation}
where
\begin{equation}
g(x) = \frac{2}{\pi} \int_0^{\pi/2} d\theta\ \cos^2\theta \exp \left(
\frac{-x\sin\theta}{2} \right) \left[ \cos \left(
\frac{x\sqrt{3}\sin\theta}{2} \right) - \frac{1}{2} \exp
\left(\frac{-x\sin\theta}{2} \right) \right] ,  
\end{equation} 
with $\theta=\pi-\varphi$. The functions $f(x)$ and $g(x)$ cannot be
represented in terms of elementary functions; however, it is not
difficult to find their asymptotic properties:
\begin{equation}
f(x) \rightarrow \left\{\begin{array}{lr} 2^{-3/2} - 2x/(3\pi) & x
\rightarrow 0 \\ 1/(\pi \sqrt{2}{x^3}) & x \rightarrow \infty
\end{array} \right.
\label{eq:asymp1}
\end{equation}
and
\begin{equation}
g(x) \rightarrow \left\{ \begin{array}{lr} 1/4 - x^2/(4\pi) & x
\rightarrow 0
\\ 3/(\pi x^3) & x \rightarrow \infty \end{array} \right.
\end{equation}
Moreover, it is readily shown from Eq.~(\ref{eq:f_func}) that
\begin{equation}
\int_0^\infty dx\ f(x) = 1/2\;.
\label{eq:asymp3}
\end{equation}
These properties are important because they allow us to obtain
$F_{1,2}(t)$ in certain regimes.

We now discuss the Coulomb case. We are interested in the thermal
contribution, which means that we approximate $\coth(\omega/2T)$ by
$2T/\omega$. Then, for $t<<t_T$, the integral in
Eq.~(\ref{eq:Fdoubleprime}) is dominated by $\sqrt{\omega}$ and we can
use the first line in Eq.~(\ref{eq:asymp1}) to get
\begin{equation}
\ddot{F}_1(t) \approx \frac{1}{2^{3/2} E_F t_T^3} \left[ 1
-\frac{2^{3/2}t}{3\pi t_T} \right] \;. 
\label{eq:f1}
\end{equation} 
For $t\gg t_T$ the convergence in Eq.~(\ref{eq:Fdoubleprime}) at high
frequencies is due to the fast decay of $f(x)$, allowing us to use
Eq.~(\ref{eq:asymp3}) and obtain
\begin{equation}
\ddot{F}_1(t) \approx \frac{1}{2 E_F t_T^2 t} \;.
\label{eq:f2}
\end{equation}
When we introduce a lower cutoff in frequency, like $\omega_c\ll T$,
there is a third regime, namely $t\gg t_{\omega_c}$, where we can
replace $f(x)$ by its large-$x$ asymptotic form. We then get
\begin{equation}
\ddot{F}_1(t) \approx \frac{1}{2^{5/2} \pi E_F} \left(
\frac{T}{\omega_c t^3} \right)\;.
\label{eq:f3}
\end{equation}
Notice that in all regimes it is legitimate to neglect the frequency
dependence coming from the $\cos(\omega t)$ factor. This means that
the explicit $t$ dependence of ${\bf a}({\bf r},t)$ given by
Eq.~(\ref{eq:phase}) can be neglected. Integrating Eqs.~(\ref{eq:f1}),
(\ref{eq:f2}), and (\ref{eq:f3}) twice over $t$ and recalling that
$F_\eta(t)\sim t^2$ for $t\rightarrow 0$, one finds
Eq.~(\ref{eq:FCoulomb}), as well as Eq.~(\ref{eq:statC}). The
calculations are analogous for the short-range case, with the
exception that in all three regimes a lower cutoff in frequency,
$\omega_c$, is required to achieve convergence.

\section{}
\label{sec:appdB}

Equation~(\ref{eq:statC}) can be derived in the following way. We
first notice that, by going to Fourier space, we can rewrite
Eq.~(\ref{eq:static}) in terms of separate frequency and spatial
integrations,
\begin{equation}
W_s = p^2 \int \frac{d^2q}{(2\pi)^2} \int_0^{\omega_c/p}
\frac{d\omega}{2\pi} \left[ \int_{R_c} d^2r\ e^{i{\bf q}\cdot {\bf r}
} \right]^2 \langle |h({\bf q},\omega)|^2 \rangle \;,
\label{eq:apbb1}
\end{equation}
where we have only kept the static part ($\omega<\omega_c/p$) of the
gauge fluctuation spectrum. The spatial integration gives $2\pi R_c
J_1(qR_c)/q$, where $R_c=v_F/\omega_c$ is the cyclotron radius and
$J_1(z)$ is the first-order Bessel function. Recalling that $\langle |
h({\bf q},\omega) |^2 \rangle = q^2 \langle | {\bf a}({\bf q},\omega)
|^2 \rangle$, we can carry out the integration over frequency,
\begin{equation}
\int_0^{\omega_c/p} \frac{d\omega}{2\pi} \coth \left(
\frac{\omega}{2T} \right) \frac{(2\pi q^3 \omega /k_F)}{ \omega^2 +
C_1^2 q^4} = \frac{\pi T q}{k_F C_1} 
\label{eq:apbb2}
\end{equation}
for $q\ll\sqrt{\omega_c/(pC_1)}$. Therefore, we have that
\begin{eqnarray}
W_s & = & \frac{2\pi^2 p^2 T R_c^2}{k_F C_1}
\int_0^{\sqrt{\omega_c/(pC_1)}} dq\ \left[ J_1(qR_c) \right]^2
\nonumber \\ & \simeq & \frac{ \pi p^2 T }{ m C_1 \omega_c} \ln \left(
\frac{E_F}{p\omega_c} \right) \;,
\end{eqnarray}
for $E_F\gg\omega_c$. 

Notice that if we remove the upper limit of the frequency integral in
Eq.~(\ref{eq:apbb1}) and treat the gauge field as purely static, we
would still obtain Eq.~(\ref{eq:apbb2}), but with the modified
restriction $q\ll\sqrt{T/C_1}$. As a result, the momentum integration
would lead instead to ($p=1$)
\begin{equation}
W_s^{\rm pure} \simeq \frac{ \pi T }{ m C_1 \omega_c} \ln \left(
\frac{T E_F}{\omega_c^2} \right) \;,
\end{equation}
which is equal to $W_s + W_d$.



\begin{figure}
\setlength{\unitlength}{1mm}
\begin{picture}(150,130)(0,0)
\put(5,10){\epsfxsize=14cm\epsfbox{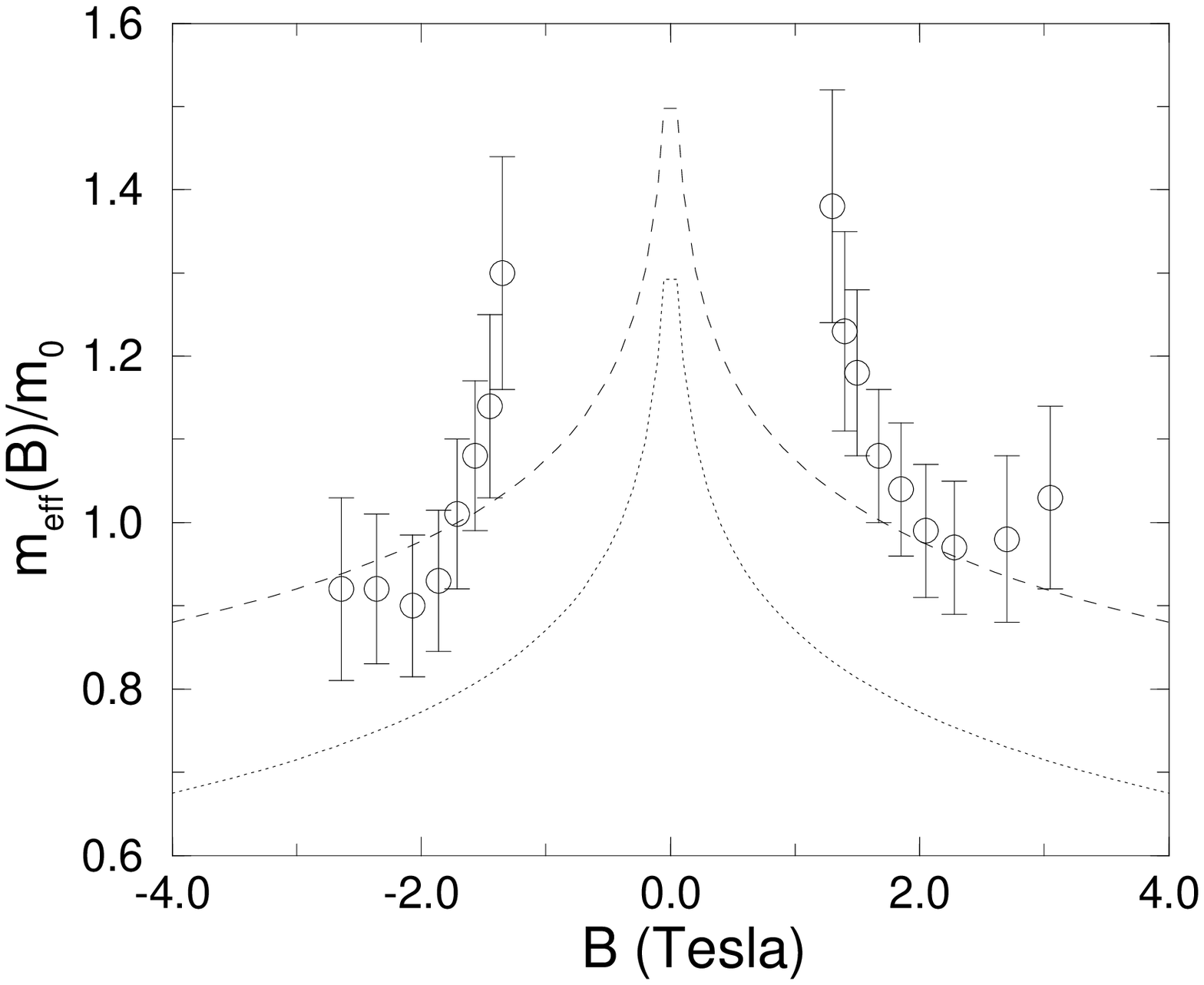}}
\end{picture}
\caption{The effective mass of composite fermions as extracted from
Shubnikov-de Haas measurements near the $\nu=1/2$ state. The circles
indicate data from Ref.\protect{\onlinecite{Du94}}. The dashed line is
a fit of Eq.~(\protect{\ref{eq:sigmaosc}}) to the $|B|\ge2$T part of
the data, resulting in $m=0.76m_0$. The dotted line corresponds to the
same expression evaluated with $m=0.55m_0.$}
\label{fig:datafit}
\end{figure}

\end{document}